\documentclass[manuscript]{acmart}

\hypersetup{
     colorlinks=true,
     linkcolor=blue,
     filecolor=blue,
     citecolor = black,      
     urlcolor=cyan,
     }
     
\usepackage{booktabs} 
\usepackage{verbatim} 
\usepackage{balance}
\usepackage{float}

\usepackage{array}
\newcolumntype{L}[1]{>{\raggedright\let\newline\\\arraybackslash\hspace{0pt}}m{#1}}
\newcolumntype{C}[1]{>{\centering\let\newline\\\arraybackslash\hspace{0pt}}m{#1}}
\newcolumntype{R}[1]{>{\raggedleft\let\newline\\\arraybackslash\hspace{0pt}}m{#1}}
\usepackage{multirow}

\usepackage{listings} 
\setcopyright{rightsretained}

\acmISBN{978-1-4503-5628-2/18/08}

\copyrightyear{2025} 
\acmYear{2025} 
\acmConference[ICER '25]{2025 International Computing Education Research Conference}{August x--x, 2025}{Virginia, USA}

\acmDOI{}
\acmISBN{}

\acmPrice{15.00}

\begin{document}
\title{SPIRAL integration of generative AI in an undergraduate creative media course: effects on self-efficacy and career outcome expectations}

\author{Troy Schotter}
\email{troy.schotter@maine.edu}
\affiliation{%
  \institution{University of Maine}
   \city{Orono}
  \state{Maine}
  \country{USA}
}

\author{Saba Kawas}
\email{skawas@uw.edu}
\affiliation{%
  \institution{University of Maine}
   \city{Orono}
  \state{Maine}
  \country{USA}
}

\author{James Prather}
\affiliation{%
  \institution{Abilene Christian University}
   \city{Abilene}
  \state{Texas}
  \country{USA}
}
\email{jrp09a@acu.edu}

\author{Juho Leinonen}
\affiliation{%
  \institution{Aalto University}
   \city{Helsinki}
  \state{}
  \country{Finland}
}
\email{juho.2.leinonen@aalto.fi}

\author{Jon Ippolito}
\email{jippolito@maine.edu}
\affiliation{%
  \institution{University of Maine}
   \city{Orono}
  \state{Maine}
  \country{USA}
}

\author{Greg L Nelson}
\email{gregory.nelson}
\affiliation{%
  \institution{University of Maine}
   \city{Orono}
  \state{Maine}
  \country{USA}
}

\renewcommand{\shortauthors}{Schotter et al}
\renewcommand{\shorttitle}{SPIRAL integration of GenAI in an undergraduate creative media course}

\begin{abstract}
Computing education and computing students are rapidly integrating generative AI, but we know relatively little about how different pedagogical strategies for intentionally integrating generative AI affect students' self-efficacy and career interests. This study investigates a SPIRAL integration of generative AI (Skills Practiced Independently, Revisited with AI Later), implemented in an introductory undergraduate creative media and technology course in Fall 2023 (n=31). Students first developed domain skills for half the semester, then revisited earlier material integrating using generative AI, with explicit instruction on how to use it critically and ethically. We contribute a mixed methods quantitative and qualitative analysis of changes in self-efficacy and career interests over time, including longitudinal qualitative interviews (n=9) and thematic analysis. 

We found positive changes in both students' creative media self-efficacy and generative AI use self-efficacy, and mixed changes for ethical generative AI use self-efficacy.
We also found students experienced demystification, transitioning from initial fear about generative AI taking over their fields and jobs, to doubting AI capability to do so and/or that society will push back against AI, through personal use of AI and observing others' use of AI vicariously. For career interests, our SPIRAL integration of generative AI use appeared to have either a neutral or positive influence on students, including widening their perceived career options, depending on their view of how AI would influence the career itself. These findings suggest that careful pedagogical sequencing can mitigate some potential negative impacts of AI, while promoting ethical and critical AI use that supports or has a neutral effect on students' career formation. To our knowledge our SPIRAL integration strategy applied to generative AI integration is novel.

\end{abstract}

%
%


\maketitle

\section{Introduction}
 
Many people in and outside of the published literature are experimenting with integrating generative AI into classes while trying to understand the impacts it has on students~\cite{prather2024beyond, denny2024CACM}. 
Using generative AI may help people learn computing, and students seem to find it helpful~\cite{vadaparty2024cs1, liu2024teaching}, but generative AI tools might increase equity gaps when more privileged learners start with more of the skills required to effectively use generative AI tools~\cite{prather2024widening}. For example, if learners don't understand the code generated by AI, using the AI could actually make them learn slower, making more mistakes and becoming more frustrated. Over-reliance on generative AI -- a concern raised by many~\cite{prather2024robots,lau2023ban,becker2023programming,prather2024beyond} -- could lead to students developing an `illusion of competence' where they do not realize they are not learning the content they should be learning~\cite{prather2024widening}. For learners from marginalized groups this could lead to them quitting earlier, particularly if more privileged learners have enough enabling skills to use generative AI to appear even farther ahead. While it has been suggested that generative AI could ``level the playing field'' between novices and experts (e.g., by some of the interviewees in a study by Lau and Guo~\cite{lau2023ban}), it is still unclear when, how, and if generative AI should be integrated into learning and how that affects learners' in a variety of ways over time. For example, a recent ITiCSE working group found that in general, the results for utilizing generative AI for code writing in computing education have been mixed~\cite{prather2024beyond}, with some studies finding positive and others negative impacts on learning.

Several instructional design strategies for deeply integrating generative AI have been explored; we are still beginning to understanding the effects of and different possible approaches. An early example is a complete CS1 curriculum integrating generative AI from the very first exercise developed by Leo Porter and Daniel Zingaro~\cite{porter2024learn}. The curriculum de-emphasizes syntax and writing code from scratch, focusing instead on utilizing available generative AI tools such as GitHub Copilot and ChatGPT to enable students to work on larger projects. Experiences from using the curriculum reported it enabling creative, open-ended projects that are larger than those students worked on in prior course iterations~\cite{vadaparty2024cs1}. In general, most educators seem to agree that the competencies required for programmers in the industry are changing due to generative AI, although a minority have yet adapted their teaching to account for this change~\cite{prather2024beyond}. For those who are adapting, they are, for example, increasing the weighting of (invigilated) exams in assessment and changing focus from assessing the final product to assessing the process~\cite{prather2024beyond}.

In integrating generative AI, learners' self-efficacy is a particularly important outcome to investigate, as it plays a central role in career 
development and other long term learning outcomes. 
Socio-cognitive career theory (SCCT) has a wide evidence base \cite{lentSocialCognitiveCareer2019} and has found that self-efficacy, career outcome expectation, and students' observation of their performance when doing learning tasks all influence choosing to continue to learn and pursue a career \cite{lentSocialCognitiveModel2013}, which is one important goal of education. Critically, self-efficacy is a primary causal path for the effect of task performance on continuing to learn \cite{lentSocialCognitiveModel2013, lentSocialCognitiveCareer2008, alshahraniUsingSocialCognitive2018}. Even if AI augments learners' ability to generate outputs and perform well on learning tasks, if learners do not develop self-efficacy, learners may experience less motivation and not engage in learning activities as much or even choose to abandon career interests \cite{lentUnifyingSocialCognitive1994, lentSocialCognitiveModel2013}. And if learners see AI as better than them, thus likely in the future to take their jobs (having low career outcome expectancy), they may switch majors 
and experience other negative effects. Within SCCT perceived competitive conditions in the labor market is also an environmental factor influencing career development 
\cite{sowaSupportingChildrensCareer2023}. Recent work has also shown that self-efficacy is a crucial determining factor in responsible use of generative AI ~\cite{margulieux2024self, prather2024widening}.

When generative AI may be better than a novice human in some domain, how do we develop learner self-efficacy in that situation? And how do we do that in a way that also supports long term career development and career outcome expectations? Answering these and related questions is an important area for research, practice, and learners.

In our work, to grow student self-efficacy while also developing the critical judgment needed for effective AI collaboration, we investigated a SPIRAL curriculum approach, where the first half of the course covered the typical
learning objectives, then we repeated going through those tasks
but integrating using generative AI with explicit instruction on how to
use generative AI critically and ethically. We designed that integration strategy for learners to both a) first build their own self-efficacy with the material, and b) build understanding and domain skills required to spot quality and errors in outputs made with generative AI, to learn to use AI in a critical way. We also integrated ethical analysis and consideration throughout the course, including in learning and writing about the effects of technology disruptions, first without AI and then with AI tools. We also aimed to support career outcome expectations by including envisioning future careers as well as a panel of professionals outside academia speaking about how generative AI was affecting their industries within the discipline. We implemented this in an introductory undergraduate creative media and technology class in Fall 2023 semester.

We asked the following research questions:

\begin{enumerate}
    \item \textbf{RQ1.1:} How did using generative AI influence learners’ self-efficacy for creative media tasks and analytical writing tasks?
    \item \textbf{RQ1.2:} How did learners’ self-efficacy for using generative AI change throughout the course?
    \item \textbf{RQ2:} How did the student's perception of how AI will affect their career field change when they were exposed to AI in the class?
    \item \textbf{RQ3:} What was the relationship between the use of generative AI in the class and career outcome expectations and interests?
\end{enumerate}

To answer RQ1.1, we did a quantitative analysis of self-efficacy surveys, with the MSLQ \cite{pintrichManualUseMotivated1991}, a validated instrument, for creative media self-efficacy, and for writing self-efficacy, we used the MAWESS, a validated academic writing self efficacy instrument \cite{bratenMeasuringMultiplesourceBased2023}. For RQ1.2, we used our own generative AI critical and ethical use self-efficacy questions due to lack of a validated instrument. For RQ2 and RQ3, we did a qualitative thematic analysis on interviews and pre-post survey question responses on career interests. We contribute an analysis of this in a creative media and technology class, which is also interesting as an alternative pathway into computing.

\section{Related Work}

\subsection{Generative AI and Self-Efficacy in Computing Education}
Generative AI is drastically changing the landscape of computing education \cite{denny2024CACM}. Early models, such as Codex, could easily solve most introductory programming problems and perform well on exams \cite{finnieansley2022robots}. By late 2023, models such as GPT-4 performed almost perfectly on introductory programs and exams \cite{prather2024robots}. Early reports of instructor adaptation to generative AI showed that some were attempting to embrace it, others were more cautious, and some were banning it entirely \cite{lau2023from}. This perception is shifting toward acceptance, though in constrained ways with guardrails for better student experience \cite{prather2024beyond}. However, student experience with generative AI coding tools is still not well understood and there are indications that it could lead to a performance gap \cite{prather2024widening}.

Self-efficacy in particular is beginning to emerge as a predictor of student success with generative AI in introductory programming \cite{margulieux2024self}. Early research conducted after the release of ChatGPT showed the possibility of generative AI tools to increase student self-efficacy \cite{yilmaz2023effect}. Students with higher self-efficacy tend to use AI less often and later in the problem solving process \cite{margulieux2024self}. Higher self-efficacy has also been linked to better outcomes with generative AI coding tools, such as GitHub Copilot \cite{prather2024widening}. Students with higher self-efficacy tend to be less distracted and face less metacognitive difficulties than their peers with lower self-efficacy \cite{prather2024widening}. For these reasons, it seems important to study the relationship between different approaches to integrating generative AI and their effects on self-efficacy.

\subsubsection{Within Creative Computing Education}
Much of the prior work in how generative AI is affecting the teaching of computing has been done in programming courses, especially introductory programming (CS1)~\cite{prather2024beyond,prather2024robots}. Compared to programming courses, much less work has been done in creative computing courses, particularly for integrating generative AI.
While some work on professional artists has done deep qualitative interviews over time \cite{rajcicDiffractiveAnalysisPromptBased2024}, we are unaware of similar studies with novice creatives.

\subsection{Socio-cognitive Career Theory}
We use Socio-cognitive Career Theory (SCCT) \cite{lentUnifyingSocialCognitive1994} in our work to analyze and describe and discuss our data and our results relative to other work, without contributing to the theory itself. SCCT states that self-efficacy and outcome expectations influence the career interests a person has, intent to take action towards career goals, their actual performance and success taking that action (such as doing a programming assignment in order to learn programming), which then influences their self-efficacy and outcome expectations, creating a feedback loop \cite{lentUnifyingSocialCognitive1994, lentSocialCognitiveCareer2019}. Environmental factors and context such as social support, vicarious experience (seeing or hearing about the experience of others), and other factors, also influence that process \cite{lentSocialCognitiveModel2013, alshahraniUsingSocialCognitive2018}. Several studies before the introduction of generative AI have applied SCCT within computing which inspired our work, for example  \cite{alshahraniUsingSocialCognitive2018,lentSocialCognitiveCareer2008}.

\subsection{Incorporating Generative AI into Learning: Effects on Self-efficacy and Career Interests}

There is some work on use and experiences with generative AI using qualitative and mixed methods in learning contexts that involve creativity (such as writing and programming \cite{padiyathInsightsSocialShaping2024, sandhausStudentReflectionsSelfInitiated2024, zhangFutureLearningLarge2024}) but not for impacts on career interests. Nor for changes in self-efficacy before and after within the specific field of creative media computing. For software engineering students qualitative interviews found potential benefits as well as ethical tensions and learning barriers \cite{choudhuriInsightsFrontlineGenAI2024}. A qualitative analysis of longitudinal survey responses found increasing student adoption for coding over time, with decreasing trusting ``code written by GenAI more than the code I write'' \cite{keuningStudentsPerceptionsUse2024}. Boguslawski et al interviewed and surveyed programming students on motivation and experiences with generative AI, then analyzed qualitatively using self-determination theory (autonomy, competence, and relatedness); they found increased self-efficacy and other motivational factors, which influence career interests and career outcome expectancy, but did not analyze effects on those \cite{boguslawskiProgrammingEducationLearner2025}.

Some work has explicitly taught users to use generative AI, reporting
learning outcomes for writing, programming, and prompt engineering for text tasks \cite{whitePromptPatternCatalog2023, wangReviewLargeVision2023, leePromptEngineeringHigher2025} but not for text-to-image tools, which we also integrated into our course. For example, 
    Woo found increased self-efficacy and improved prompt engineering skills for academic writing tasks 
    \cite{wooEffectsPromptEngineering2024}. 
    Knoth studied planning tasks, evaluating prompt quality and AI generated outputs \cite{knothAILiteracyIts2024}.
    Shibani et al. evaluated critical use of AI for a writing task with an AI use manual provided to students but not taught in class \cite{shibaniUntanglingCriticalInteraction2024}.

We are unaware of work that has studied pre-post career interest changes when incorporating generative AI in a course; prior 
interview studies with developers and instructors have reported issues with some students being discouraged from ``continuing their education due to the fear that AI will make their careers obsolete'' \cite{pratherHypeComprehensiveReview2024}. A recent literature review on junior software developers (less than 5 years experience, including students) concluded with three future research recommendations including ``How do LLM tools affect early career software developers’ first jobs? Although we found that developers are using LLMs to improve their development skills, we also found developers concerned about AI taking
their jobs in the future [5, 8, 19, 23]'' \cite{ferinoJuniorSoftwareDevelopers2025}. The closest study we are aware of is a survey of professional programmers that had used ChatGPT, with AI use frequency not related to job security threat, but higher trust in AI tools, and higher productivity, associated with higher perceived job security threat (Cohen's d=0.974, 0.843, a large effect) \cite{kuhailWillBeReplaced2024}.

\subsection{Quantitative studies on generative AI, self-efficacy, and career interests}

There are survey studies on specific factors influencing adoption, use, and perceptions of generative AI for learning, but they do not evaluate self-efficacy changes similar to our spiral generative AI integration approach. They generally find positive self-efficacy at a point in time, but without pre-post studies it is harder to interpret what their results mean. There are many survey studies of student attitudes \cite{zastudil2023generative,prather2024robots,amoozadeh2024trust,keuning2024students,oyelere2025comparative}, 
and use of generative AI \cite{houEvolvingUsageGenAI2024,prather2024robots,keuning2024students,ghimire2024coding,scholl2024novice,kazemitabaar2023studying}. Many of these studies include self-efficacy measures, though few of them have pre-post change evaluation design. Some study effects on learning motivation from using ChatGPT but not career interest changes, such as \cite{zhouImpactChatGPTLearning2023}.

\subsection{Qualitative studies on generative AI, self-efficacy, and career interests within computing and creative media}

While there are a few quantitative and qualitative studies, there are no qualitative longitudinal studies with students with generative AI integrated into the learning environment with explicit instruction. There are some studies on working professionals, such as \cite{varanasiAIRivalryCraft2025a, rajcicDiffractiveAnalysisPromptBased2024}.

There are qualitative interviews from single use session of generative AI on a programming task with students. Prather et al. did an early exploration on student's use of GitHub Copilot where they both observed and interviewed students~\cite{prather2023weird}. They found that novices struggled to use Copilot effectively, for example drifting between code suggestions they do not fully understand. Similarly, Barke et al. had participants (mostly graduate students) use Copilot and talk through their use of it~\cite{barke2023grounded}. They found that participants used Copilot mainly in two ways: for `acceleration' (working on a task where they know what they want to do) or `exploration' where the user explores the solution space. In contrast to prior work focusing on single sessions where generative AI was used, our present work focuses on longitudinal experiences over time in a course integrating teaching of how to use generative AI.
    
The closest work is Boucher et al, which observed resistance and use of generative AI in a qualitative study of game developers and artists in summer professional development camp developing mobile games \cite{boucherResistanceFutileEarly2024}; in comparison, in our study we had more novice learners and gave explicit instruction on using generative AI. The pedagogy in the camp was the director encouraging students to use generative AI, particularly for brainstorming and content production, ``but many decided not to pursue it further than initial experimentation.'' and no direct instruction is mentioned, only that two interns were assigned by the director ``to look into GAI uses.'' Many student artists resisted using generative AI tools due to mismatch with their workflow and ethical concerns, while student game programmers used it more. It was noted that ``The interns noted some uncertainty around how to effectively use GAI tools, driving much of the resistance in terms of efficacy and usability.'' and ``while artists especially noted that it is faster and more desirable to draw than to engineer a prompt— [our paper's results] still points to a lack of developed skillset in prompt design. Since the internship had such a heavy focus on producing quality work in the service of developing a publishable game, it is perhaps unsurprising that the interns did not prioritize developing these skillsets further, especially since there was an uncertain payoff to such effort.'' \cite{boucherResistanceFutileEarly2024}. Prompt engineering does improve text-to-image model outputs but was not taught explicitly \cite{wangReviewLargeVision2023, leePromptEngineeringHigher2025}

\subsection{Instructional Design for Incorporating Generative AI}

We are unaware of work proposing a spiral~\cite{brunerProcessEducation1960} integration approach where domain skills are focused on without generative AI, then those tasks are returned to with generative AI tools. There is work on AI curricula, but prior to generative AI, AI capabilities have been more narrow with fewer ways to integrate them; generating work that humans formerly could only generate was not possible before. Building on previous research, we include teaching ethical use of AI and encouraging critical thinking about AI outputs, which can be inaccurate, biased, or generate false information such as confabulations. \cite{kafaiTheoryBiasTheory2019,alvarezSociallyRelevantFocused2022, songFrameworkInclusiveAI2024, ngReviewAITeaching2023a}.

Within computing, the incorporation of generative AI into teaching has been mostly holistic~\cite{prather2024beyond}, even starting from the very first exercise~\cite{porter2024learn}, which differs from our more delayed dual mastery SPIRAL integration approach. In Porter and Zingaro's generative AI CS1 curriculum, it is integrated from the beginning \cite{porter2024learn}. Garg et al. taught structured prompt engineering for doing data analysis with Python in a ~5 hour workshop setting, without mentioning instructional design for addressing quality and confabulation issues; they found increased data analysis test scores and in a qualitative deductive analysis of semi-structured interviews found themes including increased self-efficacy but did not ask about careers \cite{gargAnalyzingRoleGenerative2024, gargImpactStructuredPromptDriven2024}.

Some general frameworks for levels of generative AI integration exist without evaluating specific designs for integration. For example, Perkins proposed levels for assessment, from no AI, to assisted idea generation and structuring, assisted editing, AI task completion with human evaluation, to using AI without having to cite what content is AI \cite{perkinsArtificialIntelligenceAssessment2024}. Some works use insightful theoretical analyses to generate guidelines and example specific activities; for example, Swindell et al. argue AI should be incorporated to promote ``humanizing action'' such as ``[AI] output is relevant to the students and shows connection specific to the context and beyond the class that could not have been achieved without human (student) guidance.'' and gives two example activities: a socratic dialogue with AI around creating a community outreach plan, and students learning about their local political landscape then writing and discussing an issue of their choice, amending their positions through students engaging with each other \cite{swindellArtificialEducationEthical2024}. There are also broader policy framework works or guidelines \cite{songFrameworkInclusiveAI2024} that, while useful, do not get into how specifically to integrate generative AI, such as supplement but don't replace human interactions and ``encouraging a balanced approach'' \cite{chanComprehensiveAIPolicy2023}. Frameworks such as LAIK do not include having human learning of tasks to build domain expertise before including using generative AI as part of doing tasks \cite{al-aliLAIKYourClassroom2024}. Additionally, while they may include teaching prompt engineering, and at some point teaching correcting generative AI, they do not in the integration approach specify consistently including both in worked examples of generative AI like our approach does \cite{al-aliLAIKYourClassroom2024}.

A few works on teaching prompt engineering itself have included some evaluation of self-efficacy changes but not careers \cite{zhangSystematicLiteratureReview2024}. For example, Mzwri et al evaluated on online asynchronous course on prompt engineering designed to also support academic English learning, with an after course survey. `Notably, 81.7\% reported an enhanced ability to apply prompt engineering techniques in SDL tasks through the use of GAI tools after completing the course. '' and improvements in self-reported academic English self-efficacy on a 1-100 scale, with the limitation that students retrospectively self-assessed their proficiency at the beginning of the course \cite{mzwriImpactPromptEngineering2025}.

Some work outside computing has integrated using generative AI deeply in a course from the beginning, but without indication of any direct instruction in prompt engineering or guided examples of how to use AI well \cite{zhangSystematicLiteratureReview2024}. For example, in a mixed methods primarily qualitative study, Wood et al integrated generative AI use into a masters level course on instructional design, where students would already have some domain expertise, and reported pre-post changes in ``comfort level with AI'' and ``awareness of ethical considerations''. However the paper does not describe any explicit instruction, and generally does not describe in detail how genAI was integrated in the design of the course beyond an experiential learning approach: \textit{``incorporates hands-on experiential learning, enabling students to directly engage with GenAI tools for planning and developing educational
materials''} and reflection during assignments where students posted their results to an AI-supported class discussion platform (the paper's example lab assignment has students use generative AI to generate assessment questions for some learning objectives, then modify them to improve them, then \textit{```Share the refined questions, modifications, and reflect on the experience''} \cite{woodEvaluatingImpactStudents2024}.) 

Generative AI for pedagogy in supporting self-regulated learning (and even self-directed learning) is a promising frontier; in our paper our dual mastery SPIRAL proposes having students first develop sufficient domain expertise without AI, in order to use AI critically, then including guided instruction of critical use of generative AI revisiting earlier material. The closest framework to our work describing integrating generative AI is HCLTF  
(Human Centered Learning and Teaching Framework)\cite{kongHumanCenteredLearningTeaching2024}, a higher level framework focused on enabling self-regulated learning in classrooms with student use of generative AI. The generative AI component in that framework is ``the generative AI domain is focused on providing technological affordances and immediate feedback. In this context, generative AI is a tool that can be used to offer individualized feedback to students.''
The example lesson plan has a different within-a-task spiral structure in using generative AI at various steps in the writing process. The paper's example 5th grade lesson plan has students do a task (revise essay version 1 to create version 2), then prompt generative AI to do the same task (revise version 1), then students compare their work to AI doing the same task, termed ``human versus machine''. 
The example lesson plan does not include explicit instruction on prompt engineering, which is also not explicitly mentioned as being reflected upon. 
The paper reports teaching teachers HCLTF and found positive changes in teacher's self efficacy for integrating generative AI but did not evaluate effects on students (having 5th grade students compare their own writing to ChatGPT might be harmful for their self-efficacy, particularly without explicitly prompting ChatGPT to write at a particular level, which is not mentioned in the paper). In our dual mastery SPIRAL approach, learners would instead write an essay themselves first and get human feedback on it. Then after doing tasks without generative AI several times, later in the spiral phase, the instructor demonstrates a worked example of using generative AI in a critical way e.g. including fixing errors and iterative prompt engineering, then learners would also use AI as an assistant in creating a new essay. In summary, with our dual mastery SPIRAL approach, students would first go through at least one full iteration of instruction, writing, and revision, without generative AI, focusing on developing their domain skills first, then later integrate generative AI.

\section{SPIRAL Instructional Design for Integrating Critical Use of Generative AI} \label{design}

\subsection{Implementation in an introductory creative media course}
The testbed for this study was the Fall 2023 [Anonymous University] course "Introduction to New Media," an exploration into the history, concepts, and practices of emerging technologies, considering the beneficial and detrimental consequences of their adoption. Course topics included having students tell stories in animation, video, and hypertext; making websites with Wordpress; creating music and interactive game avatars. The course was open to both New Media majors and non-majors.

The class was co-designed (across three people) with a senior instructor with 21 years of teaching experience, and the first and second author, building upon practices learned from faculty teaching that course, using a mix of lecture and active learning group work.
This collaborative approach allowed us to balance AI integration with established best practices in the discipline. The experienced instructor provided valuable insights into student learning patterns, potential challenges, and assessment strategies specific to the creative media context.

Particularly for creative media with artistic dimensions, there is a course design trade-off between promoting student initiative and discovery versus providing explicit instruction and guidance. The usual approach of the instructor of the course leaned more towards discovery and studio-based learning, emphasizing creative exploration and peer critique. In the AI-focused sections of the course, we also implemented a more regular structure to help students navigate these unfamiliar tools. This structure included guided prompt engineering exercises, comparative analyses of AI-generated and human-created work, and systematic reflection activities. Despite this increased structure, we maintained opportunities for creative exploration and personal expression, as these are essential components of learning in creative disciplines.

\subsection{AI Course Integration Design: Dual Mastery SPIRAL}
To integrate generative AI into the course, we designed a spiral curriculum model, where the first half of the course covered the typical learning objectives, then repeats going through those tasks while integrating using generative AI and instruction about how to use generative AI critically.

Our generative AI (GenAI) integration approach includes, for each task, giving direct guided instruction via a worked example \cite{Morrison2015}, with the example shown to students of actual AI output chosen to have some good output but also quality issues, such as confabulations, in the output. Here is the instructional design for each task:
\begin{enumerate}
    \item worked example of using GenAI with a prompt template for the task, with errors/quality issues in the output 
    \item worked example of a process for using generative AI in a critical fashion, including iterative prompt engineering
    \item in-class activities using GenAI in pairs or small groups, then reflecting on and sharing out reflections on use
    \item follow-up individual or group homework assignment using GenAI to assist with the task (one per week)
    \item weekly reflection assignment on experiences for each task/week involving what happened, their reactions, potential benefits and downsides for using GenAI, their hopes and concerns about GenAI use for their learning and society, and how to use GenAI critically and ethically
\end{enumerate}

Here is an example for the task of ``getting feedback about an assignment'' that they completed. First, the instructor presented a worked example of using generative AI using this general framework for critical use, with example AI output chosen to have some good aspects in the output but also quality issues, such as confabulations.
\begin{enumerate}
    \item Start with a good prompt.
    \item Respond in ways that work well with the prompt.
    \begin{itemize}
        \item (Different for each prompt)
    \end{itemize}
    \item Be critical.
    \begin{itemize}
        \item Change the prompt to work better.
        \item Evaluate output for correctness and bias; seek other sources (e.g., Google, textbooks, etc.).
    \end{itemize}
\end{enumerate}

The instructor would then walk through the process. For example, for "Start with a good prompt", they gave an example of a prompt template for the task. For example, for the task getting feedback on an assignment, the initial template was: 

\begin{center}
    \fbox{%
        \parbox{0.8\textwidth}{%
            \noindent As an Ethics professor, please critique the following paper. Respond using the following format:
            \begin{tabbing}
                \hspace{3cm} \= \hspace{8cm} \= \kill
                \textbf{Paper:} \> [paper will be given here] \\
                \textbf{Critique:} \> [professor critique goes here] \\
                \textbf{Improvements:} \> [how to improve goes here] \\
                \textbf{Grade:} \> [number grade goes here] \\
            \end{tabbing}
            \noindent Paper: \textless to use this prompt, paste paper here\textgreater
        }
    }
\end{center}

Then the instructor walked through a prepared example of the template and AI output. The example had actual AI output but was not live run, instead a saved output (e.g. a chat transcript) was chosen to have some good aspects in the output but also quality issues, such as confabulations. This included a review of how good the response was for the task. For example, for the task of writing an essay on technology disruption winners and losers, the quality issues included confabulated citations. The instructor would also point out quality issues in the output from a lack of including instructions or details in the prompt, such as not including a rubric for the assignment in the prompt above. Then the instructor gave an example of improving the prompt and the resulting improved output.

The students then did a class activity by filling in a provided template followed by their own prompt refinement. They then reflected on the activity. The students were then given a homework assignment to do a similar task outside of class.

\subsubsection{Course topics and sequence}
In the first half of the semester, students wrote text, coded, and created media using traditional digital tools such as Photoshop, Audacity, and p5.js; in the second half, they performed the similar tasks while also using generative AI.

\label{ClassWeekSkillTable}

Tasks performed with and without AI with a corresponding homework assignment were:

\begin{itemize}
    \item Write an essay on technology disruption and winners and losers
    \item Brainstorm ideas
    \item Design a portfolio
    \item Code a game avatar
    \item Create a composite photograph
    \item Storyboard a fictional narrative
    \item Record/generate a soundscape
\end{itemize}

Tasks performed with and without AI with no corresponding homework assignment were:
\begin{itemize}
      \item Sculpt a 3D digital object
\end{itemize}

In our implementation of the SPIRAL curriculum design, some tasks were not returned to with AI, or only done with AI assistance, for the purposes of time and/or a lack of a clear analogue. The non-AI tasks, that had no AI counterpart, were creating a stick figure animation and creating a WordPress portfolio.

AI-only tasks that did not have a non-AI equivalent were: learning with a multiple-choice question tutor and a Socratic dialogue tutor, evaluating and grading an assignment, and conducting a mock interview. As part of lecture detailed instruction about prompt engineering for different AI-generated media was given, including example prompts and prompt templates. The course also included general discussions of the impact of new technologies on society and creative professions in the past and present. Weekly AI reflection assignment surveys were graded based on participation and not quality of the reflection to avoid pressuring students to fill their response with more than what they were actually thinking.

\section{Methods} \label{method}

We collected a variety of qualitative and quantitative interview and survey data from participants. We will describe the sample group, data collection, and then describe the analysis methods.

\subsection{Participants}

Fifty one students initially enrolled in the course, 7 dropped the course before the pre-survey, and 44 students participated in the research. Of the 44 participating in the research, 31 (70\%) completed both the pre and post surveys promptly. The sampled gender diversity of the 31 students had a slight male predominance: 55\% (n = 17) identified as men and 35\% (n = 11) as women. The remaining participants (10\%, n = 3) identified as non-binary (n = 2) or chose not to disclose their gender identity. 
Students reported as intended majors or minors: new media (24), computer science (5), neither (5).
The sample of 31 students primarily consisted of early college students, with first-year (35\%, n = 11), second (32\%, n = 10),  third (23\%, n = 7) and fourth-year students (10\%, n = 3). Participants' ages ranged from 18 to 22 years, with one participant reporting an age of 30 years. The sample reflected the predominantly white population at the university, with 77\% (n = 24) of respondents self-identifying as White or Caucasian, 16\% (n = 5) identifying as members of minoritized groups, and 6\% (n = 2) preferring not to disclose. Of those minoritized groups, 2 had identified as Black or African American, 2 as Native American or Alaskan Native, and 1 as Asian.

Table 1 details the demographics survey responses for the participants that we used interview data during our qualitative analysis. ``School Lunch'' represents if the student had qualified for free or reduced lunch in high school. Cells containing a ``-'' are fields that the student elected to remain unfilled on the survey.
 \begin{table}[ht]
\begin{center}
\begin{tabular}{|l|c|c|c|c|c|c|}
\hline
\textbf{ID} & \textbf{Age} & \textbf{Racial and Ethnic Identity} & \textbf{Gender} & \textbf{Year} & \textbf{Disabilities} & \textbf{School Lunch} \\ \hline
A & 20 & White or Caucasian & Man & Second year & No & No \\ \hline
B & 18 & White or Caucasian & Woman & First year & No & No \\ \hline
C & 20 & White or Caucasian & Woman & Second year & No & Yes \\ \hline
D & 22 & \begin{tabular}{c} Native American or Alaskan Native \\ White or Caucasian \end{tabular} & Woman & Third year & Yes & No \\ \hline
E & 30 & White or Caucasian & Man & First year & No & Yes \\ \hline
F & 22 & White or Caucasian & Woman & Third year & Yes & Yes \\ \hline
G & 20 & White or Caucasian & Man & Second year & No & No \\ \hline
J & - & - & Man & Second year & No & - \\ \hline
K & 22 & White or Caucasian & Man & Third year & Yes & No \\ \hline
\end{tabular}
\end{center}
\caption{Interviewee Demographics}
\label{table:demographics_full}
\end{table}

\subsection{Data Collection}

We gave a pre-survey in week 7 at the middle of the course before generative AI was integrated into the class (students were instructed to not use generative AI in the class before this point), then a post survey in week 15. The pre- and post-survey included measures for \textbf{creative media self-efficacy} using the MSLQ self-efficacy scale \cite{pintrichManualUseMotivated1991},  \textbf{career interests} ``What are some future careers you are considering after college?'', and four questions on \textbf{self-efficacy for generative AI use} (described more in \ref{SE_qs_rationale}). Due to an error, the pre-survey had only two of the four generative AI self-efficacy questions.

Students also did weekly individual reflection assignments after each assignment in both the non-AI and AI halves of the class. For example, for the first generative AI assignment, in week 7 the questions were:
\begin{enumerate}
\item MSLQ self-efficacy scale
\item A self-efficacy scale for writing with multiple sources (the MAWSES Multiple-Source Based Academic Writing Self-Efficacy Scale) \cite{bratenMeasuringMultiplesourceBased2023},
\item The MAWSES with the instructions altered to add ``while having access to generative AI'' (“Evaluate your own ability to write academic texts while having access to generative AI tools (e.g. ChatGPT)...”)
\item Our 4 question critical and ethical self-efficacy instrument 
\item A question comparing their critical and ethical self-efficay to last week (more, about the same, or less confident)
\item Qualitative free response questions self-assessing critical and ethical use, and on their experiences with AI. 
\end{enumerate}
Week 8 to 14 were the same except the MAWSES parts were removed.
Week 2 to 6 did not include the MSLQ or our generative AI self-efficacy measure, but did include a task-specific self-efficacy question, a question comparing their self-efficacy with the previous week, and qualitative free response questions self-assessing critical and ethical use of tools they used (e.g. Google). For example, in week 2 the task was ``writing an analytic essay'', see section \ref{ClassWeekSkillTable} for the tasks. Task specific self-efficacy questions were created to be directly related to the assignment (“Rate your confidence on a scale of 1 to 10 for the statement; I can create an audio asset that could be used in a video game or other creative context.”)

The first author conducted longitudinal semi-structured qualitative interviews with 13 participants, with 8 full participants interviewed four times over the semester roughly at Week 5 or 6 (pre-AI), 8 (after AI writing assignment), 12, and 15 (end of class). All students in the course were invited to volunteer to participate in these interviews, and we accepted all students who volunteered and were 18 years of age or older. These interviews included their identities, hopes and concerns related to generative AI, generative AI use in and outside of the course, experiences and impressions from using generative AI, and career interests. Each interview ranged from 45 minutes to 1 hour and were all in-person on campus in a student lounge that was reserved for each of the interviews. Interviewees received a \$35 Amazon gift card per interview, or \$47.50 per interview if they completed all four interviews. Every interview was audio recorded (with permission) by means of a phone used only for this purpose. Each interview audio file was later transcribed using automatic speech recognition on a local machine, then manually corrected while listening to the audio file.

\subsubsection{Self-efficacy instrument rationale and design choices}
\label{SE_qs_rationale}

We used the MSLQ \cite{pintrich1991manual} as a measure of creative media self-efficacy, for the pre and post surveys, as the topic of the course was creative media. This is appropriate given that the course is designed with a breadth of content across creative media; it is a survey course and the introductory course for the creative media major. Within SCCT for our analysis this was self-efficacy for professional tasks relevant for their career interests in creative media.

We also searched for self-efficacy measures at the task level for specific assignments in the class, but only found a good fit for the technology disruption ``winners and losers'' ethical impact analysis writing task. Multiple-Source based Academic Self-Efficacy Scale (MAWSES) was chosen because it most fit the task as the task required learners to research, read, integrate and cite multiple sources. For the other tasks, a specific enough self-efficacy scale to fit the task was not found. 

For self-efficacy for critical and ethical use of generative AI, we did not find a measure for that in Fall 2023 (as no such measure for any generative AI use self-efficacy with a validity argument had yet been published); thus, we used self-efficacy theory to design questions that varied across different theoretical aspects of the construct of self-efficacy \cite{mislevySociocognitiveFoundationsEducational2018}.
As discussed in more depth in \cite{Ramalingam1998}, Bandura's theory of self-efficacy includes three dimensions: magnitude (difficulty of the task), certainty of the efficacy judgment, and generality across different situations and contexts. Our questions varied magnitude (task difficulty),  certainty, and having explicit critical and ethical elements. We did not vary situation as generative AI is so broadly applicable and to keep the number of questions low. The questions were 5 point Likert from strongly disagreee to strongly agree:
Q1: I'm confident in my ability to use generative AI so that it is helpful to my goal, Q2: I'm confident in my ability to use generative AI in a way that won't raise ethical concerns, Q3: Even for difficult tasks, I can use generative AI well, Q4: I believe I can use generative AI in a critical and ethically informed matter.

\subsection{Quantitative Analysis}

To answer RQ1.1 and RQ1.2 on changes in self-efficacy before and after the generative AI integrated part of the course, we created a registered plan for the data analysis below before analyzing the data. We then did tests for normality for the self-efficacy data. For any normal distributed data,  we did t-tests to compare pre vs. post. Otherwise, we used a paired Wilcoxon signed-rank test for pre vs. post, for the creative media self-efficacy (the MSLQ), overall generative AI use self efficacy (5 point Likert Q1: I'm confident in my ability to use generative AI so that it is helpful to my goal.), generative AI ethical use self efficacy (5 point Likert Q2: I'm confident in my ability to use generative AI in a way that won't raise ethical concerns.), self efficacy for writing (the MAWESS, and self efficacy for writing with access to generative AI (an exploratory adaptation of the MAWESS with a preamble change ``Evaluate your own ability to write academic texts \textit{while having access to generative AI tools (e.g. ChatGPT)} by rating each statement...").

\subsection{Qualitative Analysis}

To answer RQ2 and RQ3, after the class concluded, the first, second, and last author conducted a joint inductive-deductive data analysis for the interview transcripts, focusing on the pre-AI interview and the end of course interview in Week 15 as the last interview was focused on following up on initial impressions and changes from the first interview. 
In the inductive phase, three researchers independently memoed and open-coded the interview transcripts, focusing on themes related to students’ use of generative AI and its impact on their future careers, guided by Socio-Cognitive Career Theory. Researchers met to discuss and refine the emerging themes to agree on the themes.  In the deductive phase, one researcher applied these predefined themes to code all the transcripts. Throughout both phases, the researchers met regularly to ensure coding consistency, address new themes, and iteratively adjust the codes to fit the data. 

\subsection{Positionality}

This section includes positionality statements from each researcher that conducted the qualitative analysis. The first author is a white Computer Science PhD student with 2 year of computing education research experience, 5 years of higher-ed CS teaching experience. His personal belief about generative AI is that it is a useful tool that can be utilized in ways that both cause harm to students learning while also being useful at accomplishing the task the student is asking of it. His experience with generative AI at the time of the interviews primarily involved text-based generative AI via ChatGPT 4.0 and earlier and generative AI based coding assist plugin (co-pilot). His goal in researching this subject was to discover in what ways generative AI was having an effect on students so that he may better incorporate and manage it his own classrooms both in terms of learning objectives and student confidence in themselves as computer programmers. In this study he attended nearly every AI section of the course and assisted the primary professor by acting as Teaching Assistant during the in-class activities while also giving some of the lectures related to Generative AI. During the thematic analysis he had been recently hired as a computer science lecturer at [Anonymous University].

The second author is a human-centered computing professor from an ethnic minority group in the US with prior industry experience as a UX researcher in educational technologies. Her research spans HCI, accessibility, and computing education. She contributed to the scope and data analysis. At the time of the analysis, she had very limited lived experience with Gen AI tools in both work and educational contexts but has observed other colleagues interacting with these tools during co-work sessions. She came to this work with an understanding of the limitations and pitfalls of the algorithms and data used to train these generative AI models and their direct impact on perpetuating systematic racial and ethnic biases, in addition to the ethical concerns around the transparency and reliability of these tools' outputs.

The last author is a white, non-binary, and disabled computer science professor and computing education researcher. They contributed to all stages of the research from design to data analysis. Their research includes introductory programming and assessment validity, and more recently generative AI, and reflection and student agency-affirming learning in computing. At the time of the analysis they had experience using generative AI text tools, including uses to help themself and others with disabilities navigate under-resourced environments. They brought cautious optimism, empathy, and concern for the long term, deeper desirable and undesirable influences of using generative AI tools.

\section{Results} \label{results}
Our quantitative and qualitative results are below organized by research question. Statistical graphics for the quantitative self-efficacy questions RQ1.1 and 1.2 are below in Figure \ref{fig_stats}.

\begin{figure}[H]

    \centering
    \includegraphics[width=\linewidth]{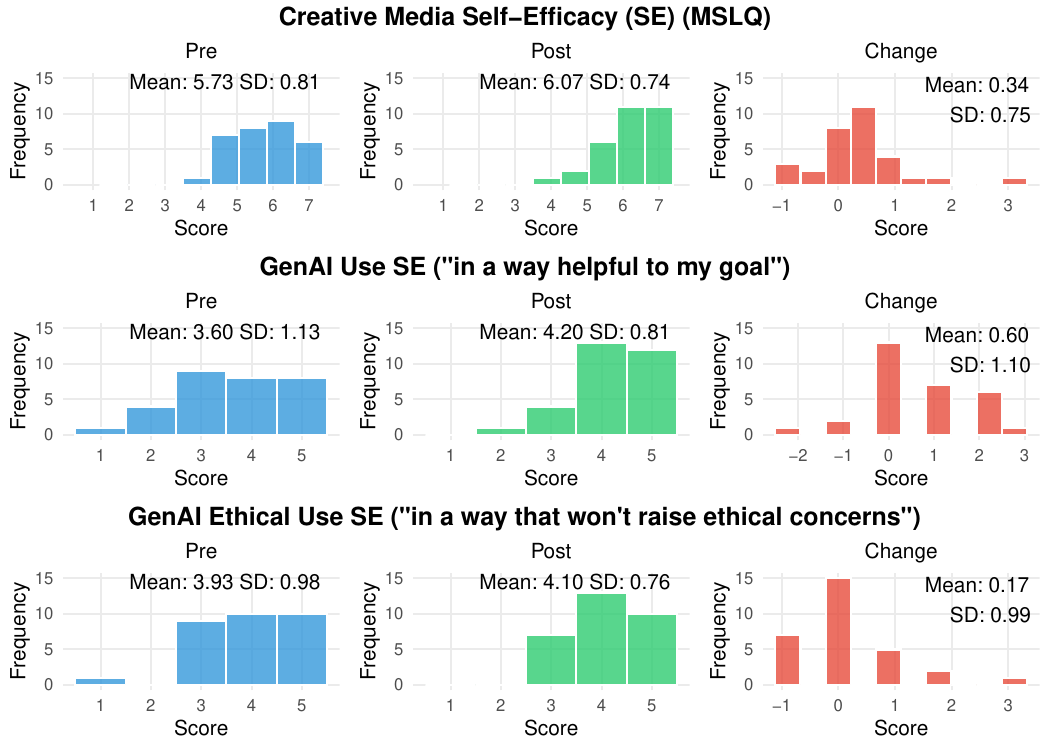}
    \caption{
    \label{fig:stats}
    Plots of the MLSQ (7 point Likert scale from 1=“Not at all true of me” to 7=“Very true of me”), and our single questions for generative AI use self-efficacy and generative AI ethical use self-efficacy (5 point Likert scale from 1=strongly disagree to 5=strongly agree).
    }
    \label{fig_stats}
\end{figure}

\begin{figure}[H]

    \centering
    \includegraphics[width=\linewidth]{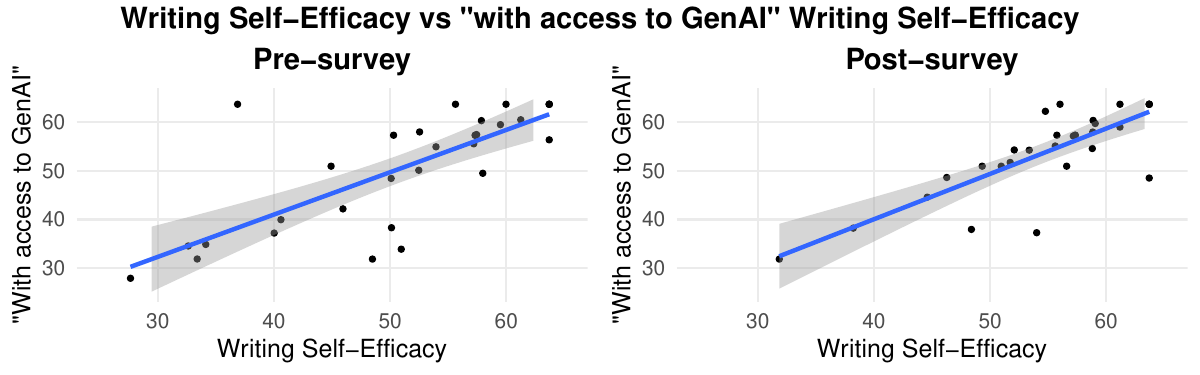}
    \caption{
    \label{fig:mawess}
    Plots of each person's MAWESS (scores made from questions from 1=``quite confident that I cannot perform this'' to 10=``quite confident that I can perform this' based on \cite{bratenMeasuringMultiplesourceBased2023}) versus MAWESS with an exploratory preamble change ``with access to generative AI, ...'', for the pre survey, then the post survey. Note also the potential change in dispersion.
    }
    \label{fig:mawess}
\end{figure}

\subsection{RQ1.1: How does using generative AI influence learners’ self-efficacy for creative media tasks and analytical writing tasks?}

Shaprio-Wilks test of normality were significant for pre and/or post data for each our quantitative measures. Thus, given our data for pre and post comparison were not normally distributed, we used Wilcoxon paired rank-sign tests for testing pre and post differences (as we specified in our registered analysis plan). We also report effect size measures \cite{ialongoUnderstandingEffectSize2016}.

For creative media self efficacy via the MSLQ, for pre vs. post, the Wilcoxon paired rank-sign test was significant, indicating a positive shift, p<.018. This indicates creative media self efficacy improved, effect size Cohen's d=.43, a small to medium effect. 

For writing self-efficacy via the MAWESS, for pre vs. post, the Wilcoxon paired rank-sign test was marginally not significant, p=.11. Writing self-efficacy did increase, Cohen's d=.31, but with our small sample size not statistically significant.

For writing self-efficacy with access to generative AI via our exploratory adaptation of the MAWESS, for pre vs. post, the Wilcoxon paired rank-sign test was not significant, p=.347. This did slightly increase, Cohen's d=.23.

\subsection{RQ1.2: How did learners’ self-efficacy for using generative AI in a critically and ethically informed way change throughout our course?}

For generative AI use self-efficacy via ``I'm confident in my ability to use generative AI so that it is helpful to my goal.'', for pre vs. post, the Wilcoxon paired rank-sign test was significant, indicating a positive shift, p<.01, with effect size Cohen's d=.63, medium effect. 

For generative AI ethical use self-efficacy via ``I'm confident in my ability to use generative AI in a way that won't raise ethical concerns.'', for pre vs. post, the Wilcoxon paired rank-sign test was not significant. Figure \ref{fig:stats} shows the change scores; this indicates a less uniform shift, and there was change up and down on the 5 point Likert scale. Overall Cohen's d=.22, small net positive effect. Note also that in learning about ethics, one might learn there are more ethical concerns than they originally thought, which may cause this to go down.

\subsection{RQ2: How did the student's perception of how AI will affect their career field change when they were exposed to AI in the class?}

We observed \textit{demystification}, where students transitioned from a more extreme perspective of what AI would do to themselves, their career, or society in general, to a more moderate view. First, they fear the unknown, AI will take jobs, moreso assuming the worst. Then, when demystified, they are less afraid because they don't think AI can take over their field. For example, for Student D in the post interview: \textit{``We fear what we don't understand and understanding it has made me fear it less.''} 

\subsubsection{Demystified from class experiences}

In the pre-AI interview students expressed more negative feelings about how generative AI would affect their world; they fear the unknown, AI will take jobs, moreso assuming the worst.

In week 5, before the AI integration, Student E was pessimistic about what generative AI would mean for their field:
\begin{quote}
 
\textit{[IBM] just held off on hiring over 5,000 positions because they hired a new AI that is going to basically just outmode all those positions, going to do their job better, unbiased, and more objectively, and they're going to pay far less per person for that AI than they will for an entire HR department, IT department... That's kind of tough. }
 
\end{quote}
After the class concluded they had lessened their fear surrounding generative AI. That AI...
\begin{quote}
\textit{
...is not just a self-evolving, unlimited potential AI that is just going to keep outgrowing, outpacing humanity until it destroys us all. I have come away from that a little bit. I realized that there are hard guardrails in place on all of these AIs that are controlled and monitored by humans, we hope, we think, we are told.  Who knows?  That is my general takeaway from this, is that it is much safer than I had originally anticipated. 
}
\end{quote}

Student F also had fears surrounding how much AI was going to invalidate their effort, and learning about generative AI was a motivating factor in the class despite their misgivings about AI. Student F had a fear of training...
\begin{quote}
    \textit{... so hard like putting all this effort in like I was talking about and then not being able to find a job and what I want to because so much of the industry has changed. That like definitely drives that part and I think that that's what motivated me to take the class even with the AI aspect. I was like if I'm gonna not like something I might as well learn about it.}
\end{quote} Afterwards, at the end of the course, F had a firm idea of where his career path could lead that would avoid the negative effects of AI:
\begin{quote}
\textit{I think I've also leaned more into not working for a company because I think companies are gonna be very productivity based. So it's gonna be more use of AI whereas if I freelance then it's like people coming to me. Like, "No I want something that's not AI generated like I like your style and I want you to create something because of that" and like that might become more niche.}
\end{quote}

Student K had a similar pathway as Student F in their first interview, 
\textit{
    ``I don't want to have a job that I enjoy and then have an AI come in and do it for me. Automate the enjoyable aspects and then so I can be cut off and not dinner that night.''
}
Like Student F, by the time the class was over he had less concerns about generative AI but still concerns about how companies would abuse it,
\textit{
    ``So as it is right now, I don't know I don't know, but I'm not phenomenally worried.  But again, I also like a lot of the creative side of things and it is very good at spewing bullshit. And again companies like, they'll put out a million shitty products over like a handful of very good ones.''
}

By the end of the course, Student B and C both developed some confidence that it was going to be people who would ultimately reject AI. Student B said: 
\textit{
``I just think I think it's always going to be limited because of the inhumanness of it the innate and the innate lack of meaning that it has.''
}
For Student C the writer's strike included an event that happened during the semester influenced their perspective: 
\textit{
I don't think people want [GenAI] to be there. I mean in the general realm of filmmaking like the writer's strike I mean that's kind of what happened when it started to come in.
...I hope that people can kind of, if it ends up coming into play in any other way that there would be a similar response.''
}

One of Student G's defining moments was a guest speaker from the field of sound design, reflected in their final interview: 
\textit{
``Especially when we had the guest speaker demo [it] sounded like a long process. So it kind of feels like that's something that an AI probably couldn't do maybe.''
}
leading to a belief that the human requirement in many jobs is safe.

Students B, E, and C had their fears lessened by trust in other people. Whether that be trusting other people to reject AI, or trusting people to place appropriate guardrails to limit AI's damage. Students F and K were more confident by the end of the class that they could carve out a niche that AI couldn't effect as much. Student G believed more that human work in jobs was more immune from AI competition.

\subsubsection{Demystification before the course} 
For students who had been exposed to generative AI earlier, their experiences also fit the demystification theme, it just had happened before the class started. J and D had an earlier demystification because they used ChatGPT and had a bad experience (J and D) or had an experience where they observed someone not learn as much because of a reliance on AI (A). 

Student J had experience using ChatGPT May 2022 when trying to have it write an appeal letter and felt the output was unacceptable for their needs. By the time the AI portion of the class was about to begin they thought of AI as something that was cool, but not as threatening (for context, J had a stutter and also reported having a learning disability) 
\textit{
``I still think that AI is cool like in terms of in terms of having it do things for you, but I feel like I feel like I feel like we shouldn't be completely reliant on AI if it malfunctions. So I think it's important for us to be able to do tasks ourselves, but I I think AI is pretty cool.''
}

Student D also had a poor experience with GenAI prior to the course where she had disappointment attempting to get AI to write an inauguration speech, saying 

\textit{
``And now, like over the last couple years...
that has, like, people have panicked about it? Kind of what? What?''
} For both J and D, additional work with generative AI did not lead to any significant change in how they felt in regards to its threat to careers.

Before our course, Student A observed his friend using AI for his coursework when A did not, and A retained more information. During a Spring 2023 creative coding course: 
\begin{quote}
    \textit{``So we're using JavaScript in P5.js Web Editor to like make art... 
    me and my friend, like we were in that class and we did like everything together. But he would always use the AI for his work and I wouldn't... 
    And I think that I retained more information by not using the AI in that class and getting like a similar grade to him.  And so yeah, like he might have like done better in the class.  But I like retained way more information from the class than he did by using ChatGPT.'' }
\end{quote}

Student A in the pre and post interview did maintain an ``it would be crazy if'' career interest for somehow getting on the ground floor at a big company that makes a big breakthrough in AI where it actually replaces many people and the company then makes a lot of money. So they had demystification but still had some vision that maybe in the future such AI developments might happen. A was also more prone to using AI in ways they self-identified as uncritical and potentially harmful for their learning.

\subsection{RQ3: What was the relationship between the use of generative AI in the class and career outcome expectation and interests?}

Generative AI use in the class had a variety of influences on the career interests and career outcome expectations of the students. We had these sub-themes:

\begin{itemize}
     \item Increased career outcome expectation from negative expectation to positive expectation, from increased self-efficacy for learning programming
     \item Increased career outcome expectation, from increased self efficacy increasing career outcome expectation, or by reducing fear AI will take over
    \item Little to no effect on career outcome expectation and field will adopt AI
    \item Little to no effect because career field will not use AI much in the future
   
\end{itemize}

For clarity, we remind the reader that this theme is \textbf{not} claiming that generative AI's existence has only had neutral or positive impacts on student career outcome expectations.  
We are only describing data reflecting a period of time from interviews given before the AI section of the course, to the end of the course when the final interviews took place.

\subsubsection{Increased self-efficacy for learning programming, changed from negative to positive career outcome expectation}

Student D had low programming self-efficacy and said she didn't have the brain for making a career out of programming, but after using GenAI to support her learning and get help, for her a programming career felt more possible (i.e. higher outcome expectancy). D identified as a Native American and white woman, and also reported having mental health conditions, anxiety and depression. 
\begin{quote}
Interviewer: Okay and this is a little bit different question how has using generative AI affected your career interests both in class, out of class?

Student D: \textbf{That's a good question that I actually have like an answer for}...
I've been coding since middle school off and on and it's like something that happened off and on because I feel like oh like it's just not for me like\textbf{ I just don't have the brain for it} but like I enjoy it. 
... \textbf{I've never thought that I could make a career out of programming} because I thought that I'd fail every job interview ... 
But with generative AI like a chatbot like ChatGPT that can help you work through your programs that you don't - I don't have to email my professor and be like we need to meet again like I'm stuck on this, like I can't debug this. Knowing that I have something to lean on when I can't figure it out and to like explain it to me step by step is like reassuring and like, like\textbf{"Oh I can totally do this" you know.} 

\end{quote}

In supporting her learning, she mentioned AI supporting her where the alternative was getting help from a professor, which is often difficult for people with social anxiety. She also found AI's not complete trustworthiness as a plus \textit{``I think that the fact that it's not completely trustworthy is kind of a positive thing in terms of tutoring because you have to actually go and check it yourself, which kind of burns it into your brain. So I think that aspect of it is a good thing. Like you can't just blindly be like, yeah, that's right, and move on...I wouldn't be like sitting with a tutor and they'd tell me an answer and I'd be like, well, let me look that up. Like that'd be so rude.''} And asked for any other differences versus a human tutor, she said \textit{``Yeah, less social anxiety around it. (laughing) Definitely like.  I personally wouldn't feel as distracted by the actual person if it was an AI. And also it's available whenever you want.''}

\subsubsection{Increased career outcome expectations}

Through an AI-driven increase of self-efficacy, some students had increased career outcome expectations.

For example, Student G reported they thought using AI will increase the quality of their work, making their potential careers more within reach, getting them to \textit{``something that I could probably figure out''}. For programming, 
\begin{quote}
\textit{It feels like a lot of the stuff's more manageable. Because like in class we, especially like when we do like the coding...
if I could get to even like a mid-tier level of coding just by using AI, then I feel like that would just open up a lot of opportunities for me.}
\end{quote}
And also for working in film, 
\begin{quote}
    \textit{If I could use AI, like having used AI to like sort of like do some basic levels of that type of thing, like editing a picture and like using it to make our own sort of sounds like a soundscape. I think that like maybe has opened me up a little bit to like, oh, this is something that I could probably figure out too.}
\end{quote}

E had a larger change opening to a new career direction in ethical AI to \textit{``make sure that it's not—that we're doing AI right by humanity, and that we're not doing AI right by companies and by governments.''} from originally 
    \textit{
    ``I was considering a career field in AI in the same exact way that somebody would put Excel Wizard on their resume when they're applying to companies that are looking for Excel Wizards. "It's a great position, but we want somebody who's very experienced with Excel."''
    }

Interaction with AI through the class increased Student A's interest in creative media as a career field: \textit{``But I'm pretty sure I am doing two [creative media] classes next semester 
I guess, like the AI, it just made me more like interested in the, like where that, the field like could go as a whole.''}

\subsubsection{Little to no effect, but their profession will adopt AI as a tool}

For these students, generative AI integration in the course had little to no effect on their career outcome expectations.
When asked about effects on their career interests, Student F had some concerns in the pre-interview about their future career, but in the post interview was more hopeful humans will still be needed in their career. 
\begin{quote}
    \textit{Even if I like want to be a marketable employee I might have to use [AI] to like cut lines.
    Like that's just kind of part of the art world in general. ...I feel like you're just gonna get tighter and tighter because it's like oh, well you can just use AI so it's gonna, it's gonna be kind of like sink or swim.}
\end{quote}
From doing the assignments in the course for illustrating a hoax and for storyboarding, F had more hope humans will still be needed in their career, \textit{``It gave me hope at how hard it was to get [AI] to do what I want...
That also gave me some hope of like, oh well you'll still be needed.
''}

For careers outside creative media, Student A also fell under this theme: \textit{
``If I were to go into New Media, I think it'd be really, I think it'd be cool if I were to do something with AI. But everything else, it doesn't, that doesn't affect my things that I've considered doing in those majors and fields...AI is likely to become more and more mainstream and important in other aspects of life.  So I will probably eventually, no matter what field, it will have some effect. But, yeah, that's just what I think in the future.''}

\subsubsection{Little to no effect, because they didn't expect their profession will use AI much in the future}

These students acknowledged AI as a threat to some careers, but not their career. This also included students potentially going into programming as a career.

Student B wants to work in either \textit{``museums, libraries, or cultural heritage institutions''} doing \textit{``archival work type stuff''} or \textit{``working in academia.''} For how generative AI has affected their career interests, B said
\begin{quote}
\textit{
...like AI can't teach college classes.  I probably could use it to potentially help in developing technologies, you know again as a tool, yeah, but I don't really, in the two minutes I'm considering it right now, don't really see how it could affect me very significantly in what I already plan to do.''}
\end{quote}

Student C is interested in working in filmmaking, and trusts that that profession or society will push AI out of their chosen career of filmmaking.
\begin{quote}
\textit{``I don't really see myself using [AI] unless, I don't want to go into a job that I have to use it so I think filmmaking for the most part stays out of that, should stay out of it...I don't think people want it to be there. I mean in the general realm of filmmaking like the writer's strike, I mean that's kind of what happened when it started to come in. So I think that I hope that people can kind of, if it ends up coming into play in any other way that there would be a similar response.''}    
\end{quote}

Student J was still deciding between multiple potential career interests as a computer science major and had also \textit{``...different media interests. I don't fully know what I want to do''}. When asked about future careers and the influence of generative AI from the class he said 
\textit{``I don't think it has, because... just... just because this is... this is AI-focused, and I don't plan to... and I don't plan to have AI matter in what I do.''} 
He also mentioned his manual work in assignments without AI \textit{``I could see myself doing those one... doing those things one day...''.}

Student K wants to be game developer or a software developer. In the post-AI interview, they described a Youtube video by a coder where ChatGPT diagnosed the cause of a bug, asked it to fix it, but ChatGPT ignored its own diagnosis:
\begin{quote}
    \textit{ 
    It's a weird.  I mean it's just a big brain with StackOverflow on it. 
...it does not work very well with like tabular data.  "Here's a table of five numbers.  Can you add them together?"  Nope. ...So as the technology is right now.  It seems like could be useful for very general like, "hey, I'm having this issue"... 
So as it is right now, I, I don't know, I don't know, but I'm not phenomenally worried. 
}
\end{quote}
So while K believed generative AI could help with some tasks, they felt like it was 

not a threat for their career.

\section{Limitations}

In the course we graded reflection surveys but only for participation. Likewise for the class assignments for the AI portion (due to ethics board conditions for our study). This may have lessened the amount of student learning due to less motivation to do higher quality work.

Our AI self efficacy measure does not have a validity argument beyond face validity. In unpublished preliminary work on a similar measure with a 7 point scale, we have found some validity evidence but it is unclear that it transfers to the 5 point scale.

Potential participant interview response bias as the interviewer was seen as an unofficial TA for the course. The interviewer had no grading power, but was present at many class periods acting in the role of a TA assisting the course.

We did not integrate into our qualitative analysis the pre and post career interest responses for non-interviewees because we lacked context to interpret them. It is a normal part of interest development to narrow interests, and sometimes to widen interests. Without richer qualitative context for the non-interviewees, with our available data we could not tell the strength and meaning of the career change. We discuss this more in future work.

Potential response bias in the surveys, including some missing data for the career interests (in SCCT, called career goals). The blanks may have been from survey exhaustion from the number of surveys in the class. When we triangulated with the interview participants who discussed their interests at length, we didn't see any pattern in non responses (we had blank responses for a person who was mostly undecided, and a person with decided career interests), but the lack of responses makes our results here more preliminary in comparison.

Our study is only one study in a particular student population, point in time, and course context; we do not claim our results generalize to other contexts. In particular, for effects on career interests and career outcome expectations, many of the students in the course had intended majors but were at a relatively undecided phase; this may have lessened generative AI threatening their career outcome expectations, since they saw themselves as having still many different potential careers. Creative media is also a discipline with inherently many potential career pathways. 

We did not do a deeper qualitative analysis of self-efficacy for the interviews. Our first paper here is an initial analysis, and while we had a couple questions about self-efficacy in the interview protocol, they did not elicit detailed responses.

\section{Discussion}

In this work, we contributed a SPIRAL (Skills Practiced Independently, Revisited with AI Later) approach to integrating generative AI, with an example implementation in an undergraduate creative media course. In the SPIRAL approach, students tackled learning objectives without generative AI in the first half of the course, and revisited them with generative AI in the second half with worked examples and practice on critical use of GenAI and iterative prompt engineering. We examined the effects of the SPIRAL generative AI integration on students' self-efficacy and career interests by conducting longitudinal interviews and surveys. Our results suggest that the SPIRAL approach increased students' self-efficacy for creative media and for using generative AI, and that students experienced \textit{demystification}, getting a more realistic understanding of the capabilities and affordances of generative AI. For career interests, the SPIRAL integration appeared to have either a neutral or positive influence on students, including widening their perceived career options, depending on their view of how AI would influence the career itself. We believe the SPIRAL approach to integrating generative AI into teaching can bring benefits while ensuring students have a more realistic understanding of GenAI capabilities and learn to use it more critically and ethically.

Our SPIRAL approach to integrating generative AI into teaching is, as far as we know, novel (for generative AI). We believe that we should measure students' ability to do tasks with and without generative AI, to see if GenAI use leads to any unintended consequences like weakening their own ability to do tasks without generative AI, which is not as easily possible in non-SPIRAL approaches, such as when it is integrated throughout the course. Recent work by Lee et al. has shown that overreliance on GenAI can weaken critical thinking skills \cite{lee2025impact}. They found that a higher confidence in GenAI is associated with less critical thinking, even though it is perceived as less effort to do so. In the domain of programming education, Prather et al. found that GenAI can directly impact student problem solving abilities, leaving them with an illusion of competence. Other recent work by Jost et al. reveal that higher use of GenAI is strongly correlated with lower final outcomes \cite{jost2024impact}. This has led some to call for measuring learning without GenAI \cite{denny2024CACM}. Our findings support this call.

Our work also provides a reason to not let students decide how to interact with AI at first. Instead, they should be provided with positive examples and frameworks of responsible use. If the first experience is poor it may solidify their perception of AI if they later go in expecting a poor experience. But once this first experience is provided, allow experimentation and discovery with an opportunity to discuss what they discover among their peers. This model follows closely with observations by Vadaparty et al. \cite{vadaparty2024cs1}.

The way that the class was formatted, with the non-AI in the first half of the semester and AI in the second half, may have played a role in how learning about GenAI affected students. In the 2024 version of this course each class topic was immediately followed by the AI equivalent of the topic in the next week. The study surrounding the 2024 version of the course did not have as good of outcomes for the students.

We observed that many uses of AI that participants judged as unhelpful or unethical, such as cheating, first occurred when they were under deadline pressure and had few other choices. We recommend strongly implementing deadline extension approaches to help students avoid transitioning from using AI unethically once into regularly using it in ways that harm their learning. For example, Flextensions automates the logistical work of managing such extensions for giving assignment extensions, and Shakir et al. argue for how extending deadlines increases learning and accessibility particularly for marginalized and disabled students \cite{shakirfuzailFlextensionsExploringImpact}.

In our instructional design we included professionals in industry talking about the impact generative AI tools and the future of professions; in our interviews we saw that reassure many students positively, and it may be more urgent now to incorporate professional panels into introductory and other courses in this time of change. At the same time, the presence of organized professional resistance to AI in some creative fields, exemplified by the writers' strike, appeared to provide students with reassurance about career viability that is notably absent in programming and other technical domains lacking similar collective action. This collective resistance seemed to bolster students' confidence that ``there are some things that humans should do'' and that creative work would remain meaningful. In contrast, programming lacks unions or similar professional organizations advocating for human roles, potentially leaving computing students more vulnerable to fears of replacement. 

Educators in fields without strong organized unions or worker communities may need to be especially attentive to addressing career anxieties when integrating AI tools. Teaching collective organizing and histories of labor movements may become increasingly an imperative over the long term, practically and to support students' career outcome expectations. In SCCT the presence or absence of visible professional communities and advocacy is an environmental factor that influences career interests and career outcome expectations. Future research should explore effects of those differences across disciplines and implications for teaching and the computing community of practice. For example, integrating student involvement in worker organizing or other community-building in their future career might be a timely activity or extra credit assignment.

In visualizing our results, we may have identified a threat to validity in measuring self-efficacy more generally in the age of generative AI, which we are unaware of in the literature. When students respond to self-efficacy questions, they may now implicitly include or not include access to generative AI tools in that judgment, and we cannot tell from their survey responses. For example, in Figure \ref{fig:mawess} there is a large change in responses, where before the AI integration students' answers are more different for writing self-efficacy versus writing self-efficacy with access to generative AI, but after they are very similar. Our study was not designed to definitively test such a change in student response processess, and we did not register an analysis plan for that as it was an unexpected observation. 
All prior validated measures for self-efficacy created before the age of generative AI may need to be re-evaluated, or even re-validated to understand what they now measure, and even be re-designed. Self-efficacy is a foundational construct in modern education theory \cite{goetzeLearningReallyJust2022, taschnerYesCanSystematic2025}, playing a role in self-regulated learning theory~\cite{artinoAcademicSelfefficacyEducational2012}, motivation \cite{huangAchievementGoalsSelfefficacy2016}, and sense of belonging and identity development \cite{trujilloConsideringRoleAffect2014, blaneyExaminingRelationshipIntroductory2017}. This may be an urgent topic for future research.

\begin{acks}
We are grateful for and acknowledge the collective human labor and effort with artificial intelligence that enabled this research, and helped editing this paper, which is currently not properly compensated and shared. 
\end{acks}

\balance
\bibliographystyle{acm}
\bibliography{sigcse} 

\end{document}